# Non-volatile electric control of spin-orbit torques in an oxide two-dimensional electron gas

Cécile Grezes[1], Aurélie Kandazoglou[1], Maxen Cosset-Cheneau[1], Luis M. Vicente Arche[2], Paul Noël[1,3], Paolo Sgarro[1], Stephane Auffret[1], Kevin Garello[1], Manuel Bibes[2], Laurent Vila[1*], and Jean-Philippe Attané[1,4*]

[1]Université Grenoble Alpes / CEA / IRIG/ SPINTEC, Grenoble, France
[2]Unité Mixte de Physique, CNRS, Thales, Université Paris-Saclay, Palaiseau, France
[3]Department of Materials, ETH Zurich, 8093 Zurich, Switzerland
[4]Institut Universitaire de France
* e-mail : jean-philippe.attane@cea.fr, laurent.vila@cea.fr

**Spin-orbit torques (SOTs) have opened a novel way to manipulate the magnetization using in-plane current, with a great potential for the development of fast and low power information technologies. It has been recently shown that two-dimensional electron gases (2DEGs) appearing at oxide interfaces provide a highly efficient spin-to-charge current interconversion. The ability to manipulate 2DEGs using gate voltages could offer a degree of freedom lacking in the classical ferromagnetic/spin Hall effect bilayers for spin-orbitronics, in which the sign and amplitude of SOTs at a given current are fixed by the stack structure. Here, we report the non-volatile electric-field control of SOTs in an oxide-based Rashba-Edelstein 2DEG. We demonstrate that the 2DEG is controlled using a back-gate electric-field, providing two remanent and switchable states, with a large resistance contrast of 1064%. The SOTs can then be controlled electrically in a non-volatile way, both in amplitude and in sign. This achievement in a 2DEG-CoFeB/MgO heterostructures with large perpendicular magnetization further validates the compatibility of oxide 2DEGs for magnetic tunnel junction integration, paving the way to the advent of electrically reconfigurable SOT MRAMS circuits, SOT oscillators, skyrmion and domain-wall-based devices, and magnonic circuits.**

The efficient control of the magnetization using currents is the key requirement to develop high performance spintronics devices for information and communication technology. In the last decade, current-induced magnetization switching by spin transfer torque has become a well-established technology, with nowadays major foundries and integrated device manufacturers commercializing Spin-Transfer Torque Magnetic Random Access Memories (STT-MRAM) [1]. Meanwhile, spin-orbit coupling emerged as an alternative method to generate spin currents and to manipulate efficiently the magnetization. SOTs rely on the injection of an in-plane current in a non-magnetic material adjacent to a ferromagnet. The spin-orbit coupling creates a spin accumulation at the interface that induces a torque on the ferromagnet magnetization. In heavy metal (HM)/ferromagnetic (FM) heterostructures [2] [3], SOTs arise from the bulk spin Hall effect (SHE) and/or the Edelstein effect associated with an interfacial Rashba state. These spin-orbit torques (SOTs) have attracted increasing attention, owing to their ability to induce magnetization oscillations or switching [4], and their potential for high speed, high endurance and low energy switching [5].

In recent years, oxide two-dimensional electron gases (2DEGs) have emerged as a new promising SOT system [6] [7] [8]. They display an efficient spin-charge interconversion through the direct and inverse Edelstein effects, arising from their broken inversion symmetry which induces a Rashba-type spin-orbit coupling. In previous works, we showed an enhancement of the spin-to-charge conversion efficiency by two orders of magnitude in such $SrTiO_3$-based 2DEG compared to conventional HM/FM heterostructures [8], along with a non-volatile electric-control of the spin-to-charge conversion [9]. In this context, the non-volatile electric-control of the reciprocal charge-to-spin conversion in these 2DEGs would be of great interest for developing reconfigurable SOT-MRAM and logic gates, offering the possibility to actively manipulate the torque by electric fields, and thus to design new architectures. The direction and amplitude of the spin accumulation is the key parameter in the design of SOT applications, as it determines the direction of magnetization switching. In conventional SOT devices based on HM/FM heterostructures, however, the relative direction between the injected current and the generated spin accumulation is fixed by the material used [10], inhibiting any dynamical reconfiguration. This feature results in two main limitations in present-day SOT-MRAMs. Firstly, the fixed direction of the SOT torque requires an inversion of polarity for the applied current in order to switch between the two magnetization states, which requires additional MRAM circuitry [11]. Secondly, the fixed direction and magnitude of the SOT torque prevents the selective switching of multiple bits sharing the same write layer in conventional SOT-MRAMs, inhibiting any in-memory logic operations. Achieving the non-volatile electric control of the SOT could thus open the way to a disruptive generation of magnetic memories and logic devices, with high-density integration and dynamical reconfigurability. Several approaches have been pursued in this direction, based on the combination of spin Hall effect heavy metal/ferromagnetic heterostructures with piezoelectric [12], ferroelectric [13] [14] and oxide materials [15] [16]

[17], or the replacement of the heavy metal by magnetically-doped topological insulators [18] (TIs). However, none of the above methods provide non-volatile control of SOT together with efficient dynamical inversion of the SOT direction in a structure compatible with magnetic tunnel junction integration.

In this work, we propose to use 2DEGs to obtain the non-volatile gate electric-field control of SOTs. An electric modulation of the 2DEG at a STO/Ta interface is achieved, providing two remanent and switchable resistivity states of the devices, with a large contrast of 1064%. We report the observation of sizable SOTs in a SrTiO$_3$/Metal system, and we further show that the SOT strength in this system can be modulated by the gate voltage with non-volatility, an inverted hysteresis being observed in the SOT dependence with the gate-electric-field. This non-volatile control of the SOT occurs over several orders of magnitude, down to full extinction and sign inversion. We also demonstrate the dynamical control of the SOT direction by the application of voltage pulses. The modulation of the SOT is found to originate from the combination of the gate dependence of the 2DEG band structure with the electric modulation of the current injection in the 2DEG.

**Magnetotransport properties**
Experiments were carried out on Ta(0.9 nm)\CoFeB(0.9 nm)\MgO(1.8 nm)\Ta(1 nm) stacks deposited onto 500 μm (001)-oriented SrTiO$_3$ substrates (cf. § Methods for details on the sample preparation), allowing the creation of a 2DEG at the SrTiO$_3$/Ta interface [19]. The deposit of a CoFeB\MgO stack on SrTiO$_3$ has been developed to take advantage of the large interfacial magnetic anisotropy of the CoFeB\MgO interface [20]. This provides perpendicular magnetization to the CoFeB ferromagnetic layer, and ensures the compatibility of such multilayers for future integration in magnetic tunnel junctions [21]. The stack was then patterned into 200 nm - 1 μm wide and 2 - 10 μm long Hall cross-bars, as shown in Fig. 1a. To characterize the magnetic and electrical properties, an input current of magnitude up to 600 μA was injected along the $x$ direction (Fig. 1b), and the longitudinal $R_s$ and transverse Hall resistances $R_H$ were measured as a function of the magnetic field, which was applied along a direction defined by the polar and azimuthal coordinates $\theta_H$ and $\varphi_H$. Hereafter, we present results obtained from 1 μm wide Hall bars, and define the injected linear current density as $j = I/w$ (where $w$ is the width of the Hall bar).

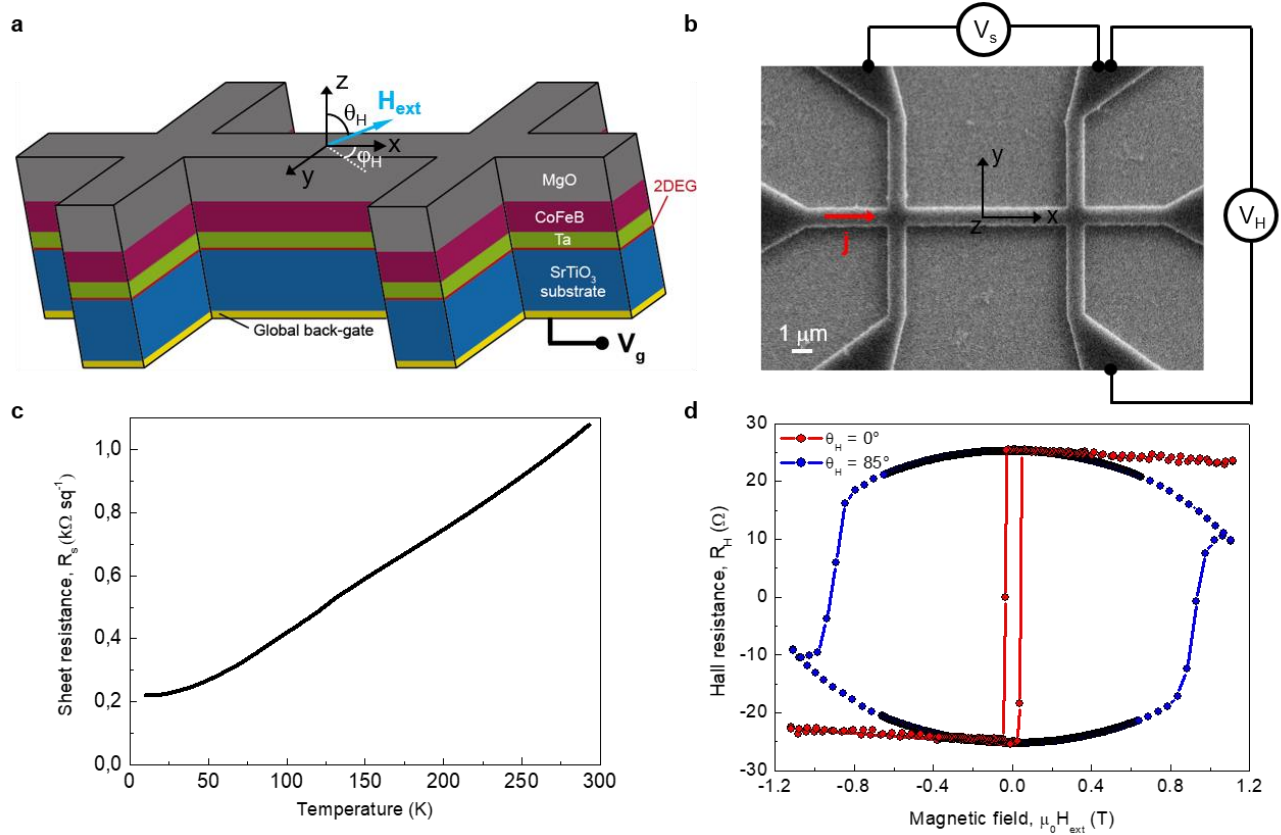

**Figure 1 | Schematic of the sample structure and its magnetotransport properties. a,** 3D schematic of the sample and of the coordinate system. **b,** Scanning electron micrograph of a Hall cross-bar and measurement principle. **c,** Temperature dependence of the stack sheet resistance $R_s$. **d,** Out-of-plane ($\theta_H = 0°$) and near in-plane ($\theta_H = 85°$, $\varphi_H = 0°$) magnetic field dependence of the Hall resistance at 10 K, showing well- defined up and down perpendicular magnetization states. All data have been measured with a 2D current density of j = 0.1 A.cm$^{-1}$.

The temperature dependence of the sheet resistance of the device is shown in Fig. 1c for j = 0.1 A.cm$^{-1}$. A reduction of the resistance by 380% is observed as the temperature decreases, which is characteristic of 2DEG conduction [8]. As shown in Fig. 1d, the anomalous Hall resistance follows a square-shaped magnetic hysteresis loop with applied out-of-plane magnetic field, indicating that the CoFeB has a perpendicular magnetization with 100 % remanence. Furthermore, a reversible decrease of the Hall resistance is observed when increasing the in-plane field, indicating a coherent rotation of the CoFeB magnetization towards the hard plane direction. From the Hall resistance $R_H = R_{AHE}\cos\theta + R_{PHE}\sin^2\theta \sin2\varphi$, where $\theta$ and $\varphi$ are the polar and azimuthal angles of the magnetization in spherical coordinates, we determine the anomalous Hall $R_{AHE} = 25.3$ Ω and planar Hall $R_{PHE} = 1.7$ Ω resistances in the ungated state (see Supplementary Section S1). Note that an anomalous Hall resistance significantly larger than that of standard SiO$_2$/Ta/CoFeB/MgO SOT devices [22] [23] is achieved in our system, taking advantage of the use of ultra-thin Tantalum layer, as previously observed by T. Zhu et al [24].

**Non-volatile electric-control of the 2DEG properties**
To modulate the 2DEG properties, an electric-field $E_g$ is applied across the STO substrate using a back-gate. Fig. 2a shows the sheet resistance $R_s$ as a function of the applied gate electric-field, at the temperature of 10 K and after initializing the ferromagnet in the up magnetization state. A hysteresis is observed in the sheet resistance, with two switchable and remanent high and low resistivity states of the device. The R$_s$ contrast, defined as (R$_{s,max}$-R$_{s,min}$)/R$_{s,min}$, shows a value of 1064%, with 615% remanent contrast at E$_g$ = 0 kV.cm$^{-1}$ between the high and low remanent resistivity states.

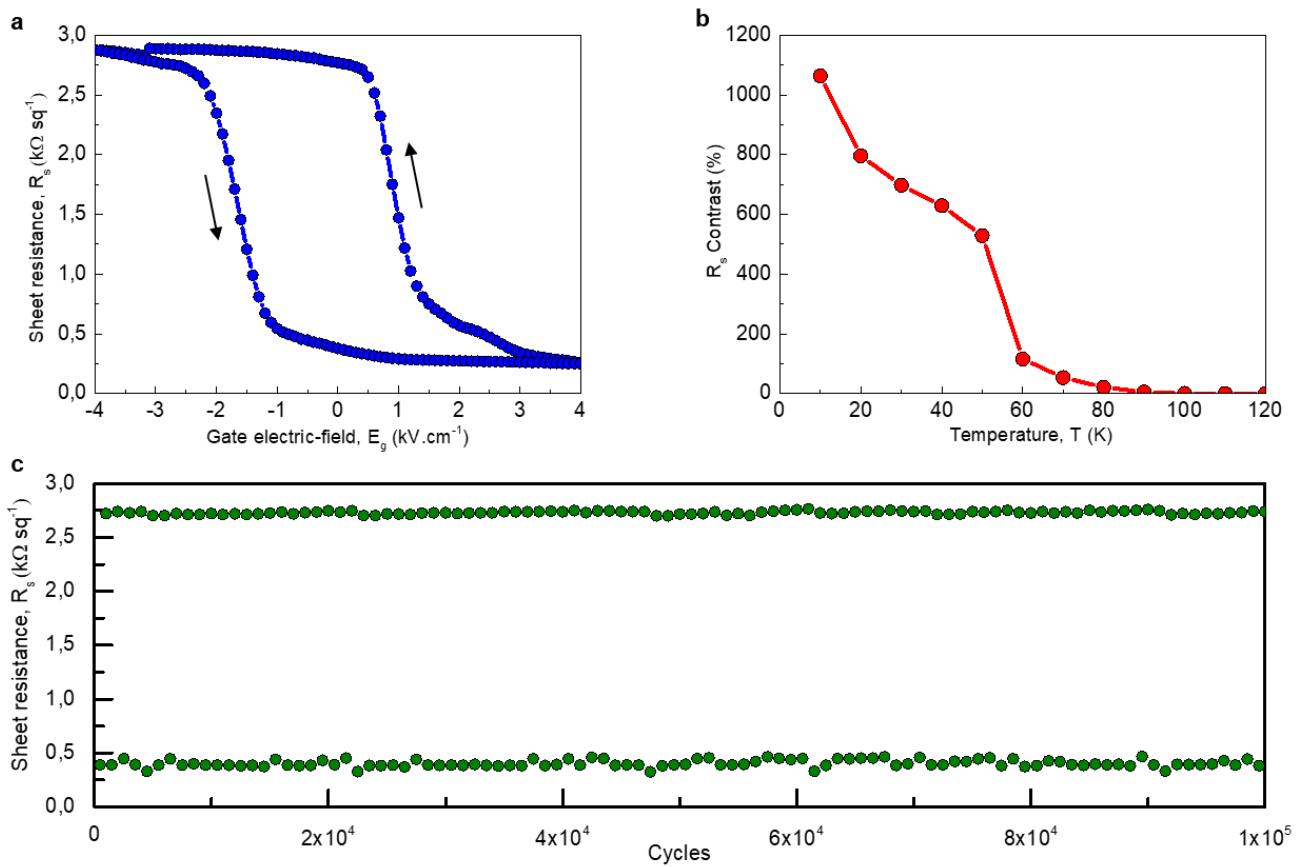

**Figure 2 | Non-volatile electric-field control of the 2D gas properties. a,** Sheet resistance R$_s$ as a function of the applied gate electric-field E$_g$ across the sample, showing high and low resistivity states of the device with R$_s$ contrast of 1064%. **b,** Temperature dependence of the R$_s$ contrast. **c,** Endurance property of R$_s$ at remanence (E$_g$ = 0 kV.cm-1) after application of positive and negative gate electric-field pulses of ±4 kV.cm$^{-1}$. All data have been measured at 10K with a 2D current density of j = 0.1 A.cm$^{-1}$, and after initialization of the ferromagnet in the up magnetization state.

The hysteresis is inverted, i.e. it is anticlockwise, which is a typical characteristic of charge trapping effects [25]. It contrasts with previous results on 2DEGs at the SrTiO$_3$\\Al interface, showing no hysteresis for 500 μm-thick SrTiO$_3$ substrates [8], and a clockwise loop compatible with field-induced ferroelectricity for thinner SrTiO$_3$\\Al substrates [9] in which larger electric fields can be applied, as well as for ferroelectric Ca:SrTiO$_3$\\Al system [26]. Here, the inverted loop rather points towards charge trapping, and leads to a larger resistance contrast. Similar anticlockwise hysteresis were observed in multiple

SrTiO$_3$\\Metal\CoFeB\MgO samples with different metal layers (see Supplementary Section S2) inducing the 2DEG formation [19]. During this process and the further annealing of the stack at 300°C, the metallic layer (tantalum) creates oxygen vacancies at the surface of the TiO$_2$-terminated SrTiO$_3$ substrate, and electron injection in the 2DEG occurs via tunneling from a few nanometers depth in the SrTiO$_3$ substrate [27]. The inverted hysteresis loop can then be understood by considering charge trapping. Large negative electric fields induce the appearance of a distribution of trapped charge near the interface. As the gate electric-field is swept from negative to positive, the 2DEG appearance is facilitated by these trapped charges [28] [29], which explains the change of resistance towards the low resistance state at negative fields. At higher fields, beyond the applied zero-gate electric-field, the electric field induces charge detrapping, leading to the symmetric behavior when sweeping the field from positive to negative values. As seen in Fig. 2b, the R$_s$ contrast decreases upon increasing the temperature until full extinction of the Rs contrast at about 105 K, corresponding to the antiferrodistortive transition from cubic to tetragonal phase of SrTiO$_3$ [30], in agreement with others charge trapping studies in SrTiO3-based 2DEG [28]. The dynamic switching between the high and low resistivity states was studied using a sequence of pulsed gate electric-field. The endurance of the device after application of successive positive and negative pulses is shown in Fig. 2c. The switching between stable high and low resistivity states shows a constant contrast, and no sign of fatigue within the 10$^5$ cycle attempts. This confirms the technological potential of the charge-trapping-related hysteretic effect observed here to control the 2DEG properties.

**Spin-orbit-torques characterization**

We used the harmonic Hall voltage measurements method to quantify the spin-orbit torques in our systems [3] [31] [32]. An AC current of frequency ω/2π = 60 Hz and amplitude I$_0$ = 400 µA (j = 4 A.cm$^{-1}$) is injected along the *x* direction to induce small oscillations of the magnetization around its equilibrium. These oscillations generate first and second-harmonic contributions to the Hall resistance R$_H$(ω) = V$_H$(ω)/I$_0$ = R$_{H,\omega}$ + R$_{H,2\omega}$, providing a sensitive way to measure the current-induced fields. The first harmonic R$_{H,\omega}$ corresponds to the Hall resistance measured in DC, while the second harmonic term R$_{H,2\omega}$ includes modulation of the Hall resistance by the SOT effective fields, R$_{SOT,2\omega}$, as well as magneto thermal effect due to unintentional Joule heating, R$_{T,2\omega}$. After subtracting longitudinal and perpendicular thermal effects contribution to R$_{H,2\omega}$, SOT effective fields can be determined by sweeping the magnetic field along the *x* (*y*) direction. In the small angles approximation, one measures a longitudinal (transverse) SOT effective fields *ΔH$_x$* (*ΔH$_y$*) respectively [31]:

$$\Delta H_{x\,(y)} = \left(\frac{dR_{SOT,2\omega}}{dH_{x\,(y)}}\right) \Big/ \left(\frac{d^2 R_{H,\omega}}{dH^2_{x\,(y)}}\right) \quad (1)$$

By defining $\xi = R_{PHE}/R_{AHE}$, the anti-damping effective field *H$_{AD}$* and field-like effective field *H$_{FL}$* are given by [31]:

$$H_{AD\,(FL)} = -2\frac{\Delta H_{x\,(y)} + 2\xi \Delta H_{y\,(x)}}{1 - 4\xi^2} \quad (2)$$

Measurements were performed with a magnetic field $\mu_0 H_{ext}$ applied close to the in-plane direction ($\theta_H$ = 85°, $\varphi_H$ = 0, 90°) and swept between ± 1.1 T. Fig. 3a,b shows the second harmonic Hall resistances $R_{SOT,2\omega}$ at $\varphi_H$ = 0° and 90°, after subtraction of the thermal effects and experimental resistance offsets (cf. Supplementary Section S3), for positive and negative gate electric-fields of ±3.2 kV.cm$^{-1}$. Typical symmetric (antisymmetric) contributions are observed for the magnetic field perpendicular (parallel) to the current, corresponding to the field like-torques and anti-damping like torques, respectively. Remarkably, we observe a sign change of $R_{SOT,2\omega}$ for high and low resistivity states. When using Eqs. (1-2), we find µ$_0$H$_{AD}$ = +6.2 mT (-0.72 mT) and µ$_0$H$_{FL}$ = +1.35 mT (-0.39 mT) for the low (high) resistivity states measured at E$_g$ = +3.2 kV.cm$^{-1}$(-3.2 kV.cm$^{-1}$), respectively. The sign change and gate dependence of SOT are discussed afterwards.

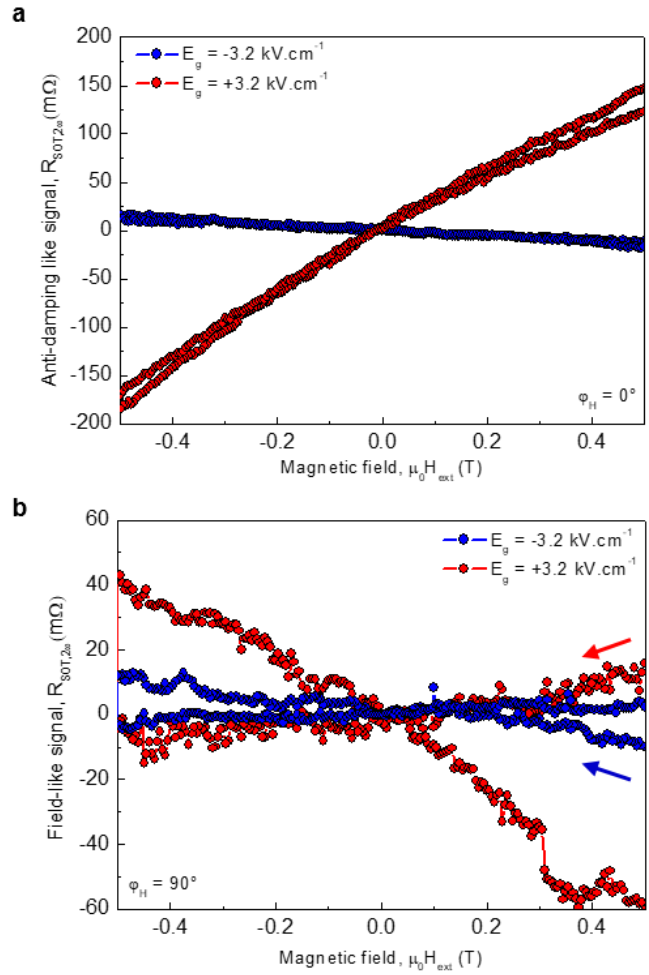

**Figure 3 | Spin-orbit torques characterization by second harmonic Hall measurement. a,** Second harmonic Hall resistances $R_{SOT,2\omega}$ at $\varphi_H = 0°$, corresponding to the anti-damping like signal, for $E_g = \pm 3.2$ kV.cm$^{-1}$. **b,** Second harmonic Hall resistances $R_{SOT,2\omega}$ at $\varphi_H = 90°$, corresponding to the field-like signal, for $E_g = \pm 3.2$ kV.cm$^{-1}$. All data have been measured at 10K with a linear AC current density of j = 4

**Non-volatile electric-field control of the spin-orbit torques**

We now repeat SOT measurements described in the previous section while modulating the 2DEG properties with gate-electric-fields ranging between ± 2.6 kV.cm$^{-1}$. Fig. 4a shows the measured SOT anti-damping like effective field per linear current density $\mu_0 H_{AD}/j$, as a function of the gate electric-field. The results show a remanent modulation of the SOT-AD effective field, with inversion of the H$_{AD}$ sign for opposite maximum gate-electric-fields. The dependence shows a hysteresis that is inverted, similarly to the resistance hysteresis shown in Fig. 2a.

Let us discuss the origin of this SOT hysteresis. The main contribution to this loop is the nonvolatile modulation of the 2DEG resistivity, and thus of the current injection in the 2DEG. Fig. 4b shows the measured modulation of the current injection in the 2DEG. The 2DEG and the Ta layer constitute two parallel channels of conduction, with two separate SOT contributions. At maximum negative electric-field (A), the device is in a high resistivity state which corresponds to the resistivity of the partially oxidized CoFeB/Ta bilayer (see Supplementary Section S4), indicating that the 2DEG is largely depleted and has an ultra-low conductance. Hence, there is no current in the 2DEG at this negative extrema. Inversely, the 2DEG resistivity reaches 0.26 kΩ.sq$^{-1}$ for the maximum positive electric-field (C). The device is then in the low resistivity state, which means that 92% of the applied current is flowing in the 2DEG.

A negative SOT-AD effective field of -0.23 mT.A.cm$^{-1}$ is observed at the negative end of the sweep (A) when there is no current flowing in the 2DEG. This suggests that the SOT is then due to a residual contribution from the Ta layer, this interpretation being in good agreement, both in sign and magnitude, with previous observations on SiO$_2$/Ta/CoFeB/MgO [22].

The highest SOT efficiency is achieved at the positive end of the electric-field sweep (C), when the 2DEG is in the low resistivity state, reaching +1.6 mT.A.cm$^{-1}$. Notably, the SOT-AD effective field at this extrema is positive, of opposite sign compared to that of the Ta/CoFeB/MgO system [22] (see Supplementary Section S4). This confirms that when the device is in the low resistivity state, the SOT arises from the 2DEG via the Edelstein effect, rather than being due to a residual Spin Hall effect in the Tantalum layer.

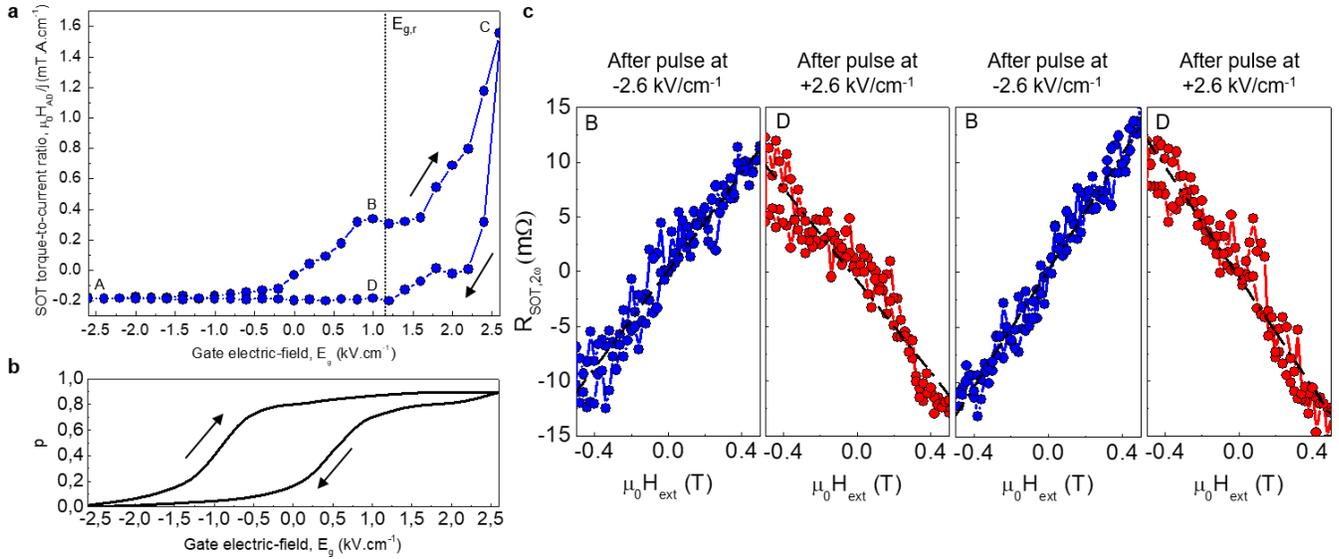

**Figure 4 | Non-volatile electric-field control of the spin-orbit torques. a,** Gate electric-field dependence of the spin-orbit torques anti-damping -like effective field $\mu_0 H_{AD}$. The points A-D are examined in **c**. **b,** Measured portion p of the current injected in the 2DEG as a function of the gate electric-field $E_g$. **c,** Normalized second harmonic Hall resistance at shifted electrical remanence $E_{g,r}$ after successive application of negative (blue) or positive (red) gate electric-field pulses of ±2.6 kV.cm$^{-1}$. Dashed and dotted lines are linear fits, yielding SOT effective fields $\mu_0 H_{AD}/j$ = +0.29, -0.20, +0.32 and -0.18 mT. A. cm$^{-1}$, respectively. All data have been measured at 10K with a linear AC current density of j = 4 A.cm$^{-1}$ and a nearly in-plane magnetic field ($\theta_H$ = 85°, $\varphi_H$ = 45°).

If the nonvolatile modulation of the 2DEG resistivity was the sole origin of the SOT hysteresis, one would expect to have similar hysteresis loops for the SOT (Fig. 4a) and for the resistance (Fig. 4b). However, the SOT hysteresis of Fig. 4a does not simply reflect the charge current redistribution within the stack. The differences between the loops of Figs. 4a and 4b are very probably due to the dependence of the spin-orbit conversion with the position of the Fermi level, which is linked to specific points of the *k* space and which has been previously studied in SrTiO$_3$\Al [8] and SrTiO$_3$\LAO [6] samples. For a given current density flowing in the 2DEG, the produced SOT is expected to vary with the applied electric field. The SOT hysteresis has thus to be understood as resulting from both the current redistribution in the stack when varying the 2DEG resistivity, and the variations of the conversion with the 2DEG Fermi level position.

The dynamical control of the SOT-AD effective field is further evidenced in Fig. 4c, which displays the measured normalized second harmonic Hall resistance at shifted electrical remanence $E_{g,r}$, after application of negative and positive 200 ms-long gate electric-fields pulses of ±2.6 kV.cm$^{-1}$. A gate electric-field pulse is first applied to the device, then the first and second harmonic Hall resistances are simultaneously measured at $E_{g,r}$. The reproducible inversion of the SOT torque sign is demonstrated, achieving $\mu_0 H_{AD}/j$ = +0.30 ± 0.1 mT.A.cm$^{-1}$ and -0.19 ± 0.1 mT.A.cm$^{-1}$ after negative and positive voltage pulses, respectively. The control of the SOT shows a deviation of less than ±0.1 mT.A.cm$^{-1}$ over successive trials.

For applications, it would be interesting to have a large remanence of the SOT at zero electric field. Here the experiment was not performed at zero electric field due to the small SOT difference induced by intrinsic electric-field dependence of the band structure in our SrTiO$_3$/Ta system, but at $E_{g,r}$, where the points B and D of Fig. 4a are well splitted. Note however that the intrinsic electric-field dependence of the band structure strongly dependent on the stack materials, as observed by L.M Vicente-Arche et al. [19] for different SrTiO$_3$/Metal system. Hence, further material engineering can be made to ensure a large contrast at zero-field remanence for applications.

**Current and temperature dependence of the SOTs**

The SOT characterizations were repeated for linear current densities ranging from 1 to 6 A.cm$^{-1}$, for the device in the high ($E_g$ = -2.6 kV.cm$^{-1}$) and low ($E_g$ = +2.6 kV.cm$^{-1}$) resistivity states. Fig. 5a shows that the amplitude of the SOT-AD effective field scales linearly with the linear current density, with a positive (negative) slope for the low (high) resistivity states, respectively. The linear scaling behavior confirms that Joule heating has a negligible impact on magnetic properties within the current range used for characterization [3]. This result further demonstrates that the SOT sign inversion is reproducible, with a difference as large as 8.7 mT at a current density of 6 A.cm$^{-1}$.

Finally, the temperature dependence of $\mu_0 H_{AD}/j$ in the high and low resistivity states is shown in Fig. 5b. In the high resistivity state ($E_g = -2.6$ kV.cm$^{-1}$), the result shows a constant AD-SOT torque-to-current ratio of $-0.23$ mT. A. cm$^{-1}$, with a less than ±0.01 mT. A. cm$^{-1}$ deviation in the 10 K to 80 K temperature range. This is in good agreement with our hypothesis of a current injected only in the Ta layer, generating a SOT which is expected to be independent of the temperature [22].

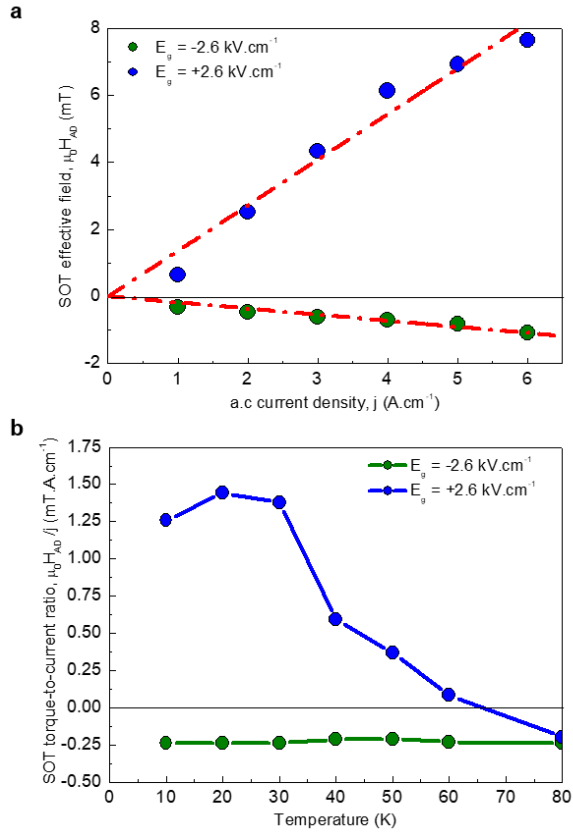

**Figure 5 | Dependence of the spin-orbit torques anti-damping like components on the injected current density and on the temperature. a,** Anti-damping like effective field $\mu_0 H_{AD}$ as a function of the applied 2D current density for positive (blue) and negative (green) gate electric-field of ± 2.6 kV.cm$^{-1}$ at the temperature of 10K. Dashed and dotted lines are linear fits, yielding SOT effective fields $\mu_0 H_{AD} = +1.36$ mT. A. cm$^{-1}$ (-0.18 mT. A. cm$^{-1}$) for positive (negative) gate electric-field. **b,** Temperature dependence of the SOT anti-damping like effective field $\mu_0 H_{AD}$ at j = 2 A. cm$^{-1}$ for positive (blue) and negative (green) gate electric-field of ± 2.6kV.cm$^{-1}$. All data have been measured with a nearly in-plane applied magnetic field ($\theta_H = 85°$, $\varphi_H = 45°$).

At the opposite gate electric-field ($E_g = -2.6$ kV.cm$^{-1}$), $\mu_0 H_{AD}/j$ is found constant from 10 K to 30 K range, at around $+1.35 \pm 0.10$ mT. A. cm$^{-1}$, followed by a decrease with increasing temperatures, reaching -0.23 mT. A. cm$^{-1}$ at 80 K. Notably, the AD-SOT torque-to-current ratio at this temperature of 80 K is identical for $E_g = \pm 2.6$ kV.cm$^{-1}$, suggesting a common origin. This result is in good agreement with the measured $R_s$ contrast displayed in Fig. 2b, in which we observed a merging of the high and low resistivity states above 80 K. Above this temperature range, the conductivity in these 2DEGs is known to disappear [19], and thus the current mostly flows in the Ta layer, which becomes the only SOT generator.

To conclude, the non-volatile electric control of the spin-orbit torques in 2DEGs could open the way to a new generation of spin-orbit torques devices. Concerning memory applications, the additional functionality provided by the electric-control can be used for building reconfigurable SOT-MRAM, in novel architectures suitable for efficient and fast operation. While efforts remains to be made to bring this technology to room temperature, the development of CoFeB/MgO heterostructures on SrTiO$_3$ with strong CoFeB perpendicular magnetization ensures the compatibility for future integration in magnetic tunnel junctions. Beyond memory applications, this non-volatile control of SOTs offers an additional way to manipulate skyrmions, domain walls or magnons by permitting a local control of the SOT along circuits. It could also be used in SOT oscillators, to allow developing logic architectures and agile terahertz emitters.

## Methods

The Ta(0.9 nm)/CoFeB(0.9 nm)/MgO(1.8 nm)/Ta(1 nm) samples have been deposited by DC magnetron sputtering on $TiO_2$-terminated (001)-oriented STO substrates of 500 μm thickness (from SurfaceNet). $TiO_2$-termination was achieved through a chemical treatment, where the substrate was subsequently submerged in $H_2O$ for 10 minutes and an acid solution (HCl 3 : $HNO_3$ 1 : $H_2O$ 16) for 20 minutes. The Hall bar devices were patterned by electron-beam lithography into 200-1000 nm-wide crosses, and subsequent deposition and lift off of the stack. After deposition, the stack was annealed at 300°C during 10 minutes to crystallize the MgO. A sample-wide back-gate of Ti(10 nm)/Au(100 nm) was then added by evaporation. The Hall voltage measurements were performed by using an AC current with an amplitude of 200 to 600 μA, modulated at $\omega/2\pi$ = 60 Hz. $V_{H,\omega}$ and $V_{H,2\omega}$ were recorded simultaneously using 2 lock-ins during sweeps of the external magnetic field for 6 s at each field step.

## Acknowledgements

This work received support from the ERC Advanced grant number 833973 "FRESCO", the H2020 ITN project SPEAR, the French Research Agency (ANR) as part of the projects OISO (ANR-17-CE24-0026-03) and CONTRABASS (ANR-20-CE24-0023) and in the framework of the "Investissements d'avenir" program (ANR-15-IDEX-02). The authors would like to thank the Institut Universitaire de France, the Plateforme Technologique Amont (PTA) for technical support, as well as members of Spintec and UMR Thales for fruitful discussions.

## Author contributions

C.G., L.V., J.P.A. and M.B. planned the experiment. C.G., L.V and S.A. fabricated the samples. C.G and A.K performed the measurements. C.G., K.G., L.V. and J.P.A. analyzed the data. C.G., K.G., L.V. and J.P.A wrote the manuscript. All authors discussed the results and the manuscript.